Neutron Operators (26/IV/2015)

# Theory of neutron scattering by electrons in magnetic materials

S. W. Lovesey[1,2]

1. ISIS Facility, STFC, Oxfordshire OX11 0QX, UK

2. Diamond Light Source Ltd, Oxfordshire OX11 0DE, UK

**Abstract** A theory of neutron scattering by magnetic materials is reviewed with emphasis on the use of electronic multipoles that have universal appeal, because they are amenable to calculation and appear in theories of many other experimental techniques. The conventional theory of magnetic neutron scattering, which dates back to Schwinger (1937) and Trammell (1953), yields an approximation for the scattering amplitude in terms of magnetic dipoles formed with the spin (**S**) and orbital angular momentum (**L**) of valence electrons. The so-called dipole-approximation has been widely adopted by researchers during the past few decades that has seen neutron scattering develop to its present status as the method of choice for investigations of magnetic structure and excitations. Looking beyond the dipole-approximation, however, reveals a wealth of additional information about electronic degrees of freedom conveniently encapsulated in magnetic multipoles. In this language, the dipole-approximation retains electronic axial dipoles, **S** and **L**. At the same level of approximation are polar dipoles - called anapoles or toroidal dipoles - allowed in the absence of a centre of inversion symmetry. Anapoles are examples of magneto-electric multipoles, time-odd and parity-odd irreducible tensors, that have come to the fore as signatures of electronic complexity in materials.

**Prologue**

The interaction between neutrons and electrons was explored in the 1930s, followed by a definitive study in 1953 by George Trammell that later was reformulated in a more compact format. Why re-visit these calculations after six decades during which time magnetic neutron scattering has become the method of choice for determining motifs of magnetic dipoles and their excitations, e.g., spin-waves? The principal motivation is to complete available calculations by including electronic operators that change sign when space coordinates are inverted, whereas the magnetic dipole is unchanged by inversion.

By way of orientation, consider the amplitude for the magnetic scattering of neutrons by electrons in the limit of small scattering angles, i.e., a small scattering wavevector $k$. The scattering amplitude is $\mathbf{Q}_\perp = [\mathbf{k} \times (\mathbf{Q} \times \mathbf{k})]/k^2$, in which an intermediate operator,

$$\mathbf{Q} \approx (1/2)\{2\mathbf{S} + \mathbf{L}\} + i\,\{\mathbf{k} \times \mathbf{D}\}/k,$$

can be justified for small $k$. The first contribution to **Q** is the magnetic dipole moment; the magnetic axial vector $\mathbf{Q} \approx (1/2)\{2\mathbf{S} + \mathbf{L}\}$ is likely to be a familiar approximation to all who use the neutron scattering technique to study magnetic materials (an atomic form factor in **Q** is approximated by unity for the moment). By contrast, the second contribution $\{i\,\mathbf{k} \times \mathbf{D}\}$ with **D** both magnetic (time-odd) and polar (parity-odd) is most likely not expected. The dipole **D** has no matrix elements different from zero if magnetic ions that contribute to it occupy sites that are centres of inversion symmetry. In the absence of a centre of symmetry a matrix element such as $\langle 3d|\,\mathbf{D}\,|4p\rangle$ for a 3d-transition ion can be different from zero. Many magnetic materials use sites that are deprived of a centre of inversion

symmetry, and it is to be noted that the crystal structure can be centro-symmetric, e.g., chromium ions in $Cr_2O_3$ with the corundum structure $D_{3d}$ ($\overline{3}$m).

In addition to completing a theory of magnetic neutron scattering by including parity-odd operators, like **D**, in this communication we draw attention to the use of atomic multipoles. These electronic entities have universal appeal, because they arise with other techniques, e.g., μSR, NMR, resonant x-ray Bragg diffraction and the Mössbauer effect.

By and large, we do not provide derivations of results. The algebra is in the literature already and to include it again, no matter how elegant the mathematics, will divert attention from results that influence the way we interpret magnetic neutron scattering data. However, we strive to make the presentation self-contained to facilitate future applications to specific experiments.

Section 1 contains essential elements of what we are about, including a complete version of that mentioned here by way of orientation. Definitions gathered in §2 are necessary equipment by which to make simulations of scattering amplitudes using an atomic picture. Sections 3 and 4 are devoted to the two contributions to the scattering amplitude, one involves electron spin and the other does not. The latter, purely orbital contribution, is technically the most demanding, because it includes the gradient operator acting on position variables. The dipole approximation for parity-even events in neutron scattering is the first port-of-call in most simulations and an appropriate expression is provided in §5. Universal expressions for the scattering amplitude for parity-even and parity-odd events are recorded in §6. Multipoles retained in expressions (6.2) – (6.7) likely provide a sound basis for most purposes. The formulation adopted for the universal expressions has proved to be efficient in many examples of x-ray and neutron diffraction. Parity-even scattering events from equivalent atomic electrons have been studied extensively in the past, and a summary of results is given in §7. Multipoles for nd- and nf-configurations, specified by fractional parentage coefficients, can be assembled from entries in tables that we provide. Many aspects of the material we review in this communication are illustrated in a worked example set out in §8. The chosen compound is an underdoped cuprate in its so-called pseudo-gap phase. Section 9 is given over to a short discussion of results and findings for neutron scattering by magnetic materials.

**1**. **Magnetic scattering operator and multipoles**

We continue with proper definitions of quantities mentioned in the prologue. The amplitude for magnetic scattering of neutrons is written $\mathbf{Q}_\perp = [\boldsymbol{\kappa} \times (\mathbf{Q} \times \boldsymbol{\kappa})]$ using a unit vector $\boldsymbol{\kappa} = \mathbf{k}/k$ where **k** is the scattering wavevector [1, 2, 3]. An intermediate operator can be written,

$$\mathbf{Q} = \exp(i\mathbf{R}_j \cdot \mathbf{k}) \, [\mathbf{s}_j - (i/\hbar k)(\boldsymbol{\kappa} \times \mathbf{p}_j)], \tag{1.1}$$

in which $\mathbf{R}_j$, $\mathbf{s}_j$ and $\mathbf{p}_j$ are operators for electron position, spin and linear momentum, respectively. Note that **Q** is arbitrary to within any function proportional to $\boldsymbol{\kappa}$.

A standard dipole-approximation for **Q** is a linear combination of the spin and orbital angular moment of the magnetic ion, **S** and **L**, respectively. The time-average, or expectation value, of the amplitude is observed by Bragg diffraction, and the averaging process is denoted by enclosing an operator by angular brackets ⟨ ... ⟩. The dipole-approximation for Bragg diffraction is thus,

$$\langle \mathbf{Q} \rangle^{(+)} \approx (1/2) \, \langle j_0(k) \rangle \, \langle 2\mathbf{S} + \mathbf{L} \rangle, \tag{1.2}$$

where the atomic form factor, a radial integral, is defined such that $\langle j_0(0)\rangle = 1$. The electronic entity in (1.2) is the magnetic moment of an ion that is a magnetic axial vector.

The superscript $^{(+)}$ in (1.2) denotes that the intermediate operator is constructed with electronic operators that are parity-even, specifically an axial dipole. If the environment of a magnetic ion is not a centre of inversion symmetry its electrons may form magnetic polar (parity-odd) dipoles, and like multipoles of higher rank. Such time-odd and parity-odd atomic entities are called magneto-electric multipoles - the name signifies the fact that fundamental operators can be formed by products of the magnetic dipole and the electric dipole, **n**, as we shall see. The spin and linear momentum operators in (1.1) both contribute magneto-electric dipoles, and the contribution to be added to (1.2) is,

$$\langle \mathbf{Q}\rangle^{(-)} \approx i\,\boldsymbol{\kappa}\times\langle \mathbf{D}\rangle, \qquad (1.3)$$

where the magnetic polar dipole,

$$\langle \mathbf{D}\rangle = (1/2)[\,i(g_1)\langle \mathbf{n}\rangle + 3(h_1)\langle \mathbf{S}\times\mathbf{n}\rangle - (j_0)\langle \boldsymbol{\Omega}\rangle]. \qquad (1.4)$$

Dipoles $\mathbf{S}\times\mathbf{n}$ and $\boldsymbol{\Omega}$ are Hermitian anapoles, with $\langle \mathbf{S}\times\mathbf{n}\rangle$ coming from $\{\exp(i\mathbf{R}_j\cdot\mathbf{k})\,\mathbf{s}_j\}$ and $\langle \boldsymbol{\Omega}\rangle$ coming from $\{\exp(i\mathbf{R}_j\cdot\mathbf{k})(i/\hbar k)(\boldsymbol{\kappa}\times\mathbf{p}_j)\}$. Cartesian components of Hermitian dipoles are purely real, which means that a Cartesian component of $\langle\mathbf{D}\rangle$ is complex when $\langle\mathbf{n}\rangle$ is different from zero. The anapoles are time-odd, and $i\mathbf{n}$ has a like property. Radial integrals $(g_1)$, $(h_1)$, $(j_0)$, and $\langle j_0(k)\rangle$ in (1.2) and (1.4) are displayed in the Figure 1.

Multipoles we have encountered encapsulate degrees of freedom in the ground-state of valence electrons, and they are defined by us to possess discrete symmetries. A magnetic multipole is time-odd, and only multipoles of this type are visible in magnetic neutron scattering. The magnetic dipole $\langle \mathbf{L} + 2\mathbf{S}\rangle$ is time-odd and parity-even (axial). A multipole of rank $K$ with the same discrete symmetries is denoted $\langle T^K_Q\rangle$, and the $2K + 1$ projections obey $-K \leq Q \leq K$. Magneto-electric multipoles are time-odd and parity-odd. Magneto-electric dipoles ($K = 1$) include a spin anapole that appears in (1.4) and the magnetic field distribution is reproduced by a solenoid deformed into a torus – the names toroidal dipole and anapole are synonymous. Magneto-electric multipoles constructed from $\mathbf{S}$ and $\mathbf{n}$ are denoted by $\langle H^K_Q\rangle$, and corresponding multipoles constructed from orbital variables alone are denoted by $\langle O^K_Q\rangle$. An actual magnetic charge, or magnetic monopole, $\langle H^0_0\rangle \propto \langle \mathbf{S}\cdot\mathbf{n}\rangle$ is not visible in neutron scattering. However, $\langle \mathbf{S}\cdot\mathbf{n}\rangle$ does contribute to the resonant diffraction of x-rays. Orbital variables do not form magnetic charge, because $\mathbf{L}$ and $\mathbf{n}$ are orthogonal, $(\mathbf{L}\cdot\mathbf{n}) = (\mathbf{n}\cdot\mathbf{L}) = 0$, while an orbital anapole can be created from $\boldsymbol{\Omega} = \mathbf{L}\times\mathbf{n} - \mathbf{n}\times\mathbf{L}$. The label magnetic charge is justified by the observation that a "magnetic charge" inserted in Maxwell's equations, with symmetries of the electric and magnetic field unchanged, is both time-odd and parity-odd. While magnetic charge in question and Dirac's magnetic monopole share the same discrete symmetries one is attached to a magnetic ion while Dirac's magnetic monopole - yet to be observed - is a fundamental unit of charge just like its brother the electron is the fundamental unit of electric charge.

Our spherical multipoles are irreducible. Thus, a quadrupole does not possess a charge or a dipole in its far-field.

For scattering by crystalline materials one needs a unit-cell structure factor created with the sum of (1.2) and (1.3) evaluated for each magnetic ion in the cell, together with spatial phase-factors determined by elements of translation symmetry in the magnetic space-group. Environments used by ions are related by elements of symmetry (inversion, rotation and improper- rotation) in most space-groups.

## 2. Definitions

In order to make our presentation self-contained and thus useful in future applications it is necessary to list certain definitions. These concern properties of spherical tensors and their signatures with respect to reversal of space and time co-ordinates. Some definitions might appear trivial at first sight but experience shows otherwise, e.g., we choose to order spin and orbital variables as s-$l$ while other workers use the reverse order, $l$-s, and this may make a difference at intermediate stages of the calculation of an observable quantity. Much material in subsequent sections can be understood and used without recourse to any of the following definitions, however.

A normalized spherical harmonic $C^a_\alpha(\kappa)$ is defined in the notation adopted by Racah, namely [4],

$$C^a_\alpha(\kappa) = [(4\pi)/(2a + 1)]^{1/2} Y^a_\alpha(\kappa), \tag{2.1}$$

where $a$ is the rank and projections $\alpha$ satisfy $-a \leq \alpha \leq a$. The complex conjugate satisfies $[C^a_\alpha(\kappa)]^* = (-1)^\alpha C^a_{-\alpha}(\kappa)$. A similar relation holds for a Hermitian multipole $\langle U^{K'}_{Q'}\rangle = (-1)^{Q'} \langle U^{K'}_{-Q'}\rangle^*$.

A spherical tensor of rank $K$ is constructed from the product of tensors $A^a$ and $B^b$ using,

$$\{A^a \otimes B^b\}^K_Q = \sum_{\alpha\beta} A^a_\alpha B^b_\beta (a\alpha b\beta | KQ)$$

$$= (1/2)\sum_{\alpha\beta} \{[A^a_\alpha B^b_\beta + B^b_\beta A^a_\alpha] + [A^a_\alpha B^b_\beta - B^b_\beta A^a_\alpha]\} (a\alpha b\beta | KQ). \tag{2.2}$$

In the second equality the product of tensors is a sum of Hermitian and anti-Hermitian operators. The latter is zero if $A^a$ and $B^b$ commute, but $\{A^a \otimes B^b\}^K_Q$ is not Hermtian even in this case. The Clebsch-Gordan coefficient in (2.2) and Wigner 3-j symbol are related by,

$$(a\alpha b\beta | KQ) = (-1)^{-a+b-Q} \sqrt{(2K+1)} \begin{pmatrix} a & b & K \\ \alpha & \beta & -Q \end{pmatrix}. \tag{2.3}$$

Two tensor products created from dipoles are $\{A^1 \otimes B^1\}^0 = -(1/\sqrt{3}) (\mathbf{A} \cdot \mathbf{B})$ and $\{A^1 \otimes B^1\}^1 = (i/\sqrt{2}) (\mathbf{A} \times \mathbf{B})$. Cartesian components of a dipole are $B_x = (1/\sqrt{2})(B_{-1} - B_{+1})$, $B_y = (i/\sqrt{2})(B_{-1} + B_{+1})$, $B_z = B_0$.

A matrix element of a spherical tensor-operator obeys the Wigner-Eckart Theorem [4]. Denoting such an operator by $V^{K'}_{Q'}$,

$$\langle JMsl| V^{K'}_{Q'} |J'M's'l'\rangle = (-1)^{J-M} (Jsl\|V^{K'}\|J's'l') \begin{pmatrix} J & K' & J' \\ -M & Q' & M' \end{pmatrix}, \tag{2.4}$$

in which $(Jsl\|V^{K'}\|J's'l')$ is a so-called reduced matrix-element (RME), and total angular momentum $\mathbf{J} = \mathbf{s} + \mathbf{l}$. Note that $(J - M)$ is always an integer. If $V^{K'}$ is a function of orbital operators alone,

$$\langle lm| V^{K'}_{Q'} |l'm'\rangle = (-1)^{l-m} (l\|V^{K'}\|l') \begin{pmatrix} l & K' & l' \\ -m & Q' & m' \end{pmatrix}. \tag{2.5}$$

An RME of an operator with defined discrete symmetries obeys two fundamental identities [6] that apply for both $J$ integer and $J$ half-integer states,

$$(Jsl\|U^{K'}\|J's'l') = (-1)^{J'-J} (J's'l'\|U^{K'}\|Jsl)^*, \tag{2.6a}$$

$$(J's'l'\|U^{K'}\|Jsl) = (-1)^{J-J'} \sigma_\theta \sigma_\pi (-1)^{K'} (Jsl\|U^{K'}\|J's'l'). \qquad (2.6b)$$

The first identity holds for an Hermitian operator, while the second identity is independent of the specific operator, because it depends solely on definitions of time-reversed states and parity. In (2.6b) $\sigma_\theta = \pm 1$ ($\sigma_\pi = \pm 1$) is the time-signature (parity-signature) of $U^{K'}$. In most cases of interest, an RME is either purely real or purely imaginary, in which case (2.6) tells us that $[\sigma_\theta \sigma_\pi (-1)^{K'}] = \pm 1$, where the upper sign applies to purely real and the lower sign to a purely imaginary RME of an Hermitian operator.

The RME of a tensor product (2.2) formed by spin and spatial variables $z^a$ and $y^b$, respectively, is usefully written in terms of a unit tensor [6]. We define such an RME as,

$$(\theta\|\{z^a \otimes y^b\}^{K'}\|\theta') = (s\|z^a\|s)(l\|y^b\|l') W^{(a,b)K'}(\theta, \theta'), \qquad (2.7)$$

where $W^{(a,b)K'}(\theta, \theta')$ is a unit tensor and composite labels $\theta = Jsl$ and $\theta' = J's'l'$. A tensor product is generally not an Hermitian operator, even when both parent operators, $z^a$ and $y^b$, are Hermitian. An Hermitian operator can be constructed from a tensor product, however, with a little ingenuity. For commuting operators introduction of a complex phase factor suffices, as in (4.2) and (7.4). Result (2.7) applies to equivalent electrons in an atomic shell, $l = l'$, discussed in §7. Configurations of equivalent electrons are usually constructed with fractional parentage coefficients - rather than Slater determinants - that appear in $W^{(a,b)K'}(\theta, \theta')$ and not the RMEs in (2.7). Estimates of unit tensors $W^{(a,b)K'}(\theta, \theta')$ are available from simulations of electronic structures [7, 8]. Parity-even and parity-odd multipoles are estimated in a package developed by Y Joly for the analysis of x-ray absorption and diffraction measurements [9].

For one electron,

$$W^{(a,b)K'}(\theta, \theta') = [(2j+1)(2K'+1)(2j'+1)]^{1/2} \begin{Bmatrix} s & s & a \\ l & l' & b \\ j & j' & K' \end{Bmatrix}, \qquad (2.8)$$

in which $s = 1/2$ that yields $a = 0$ or 1, with $(s\|z^0\|s) = \sqrt{2}$ and $(s\|z^1\|s) = \sqrt{(3/2)}$ in (2.7). The magnitude of the 9j-symbol is unchanged by an even or odd exchange of columns or rows, but an odd exchange changes its sign by a factor $(-1)^\Re$ with $\Re = (1 + a + l + l' + b + j + j' + K')$ [4, 5, 6].

Let us explore the physical content of the two identities (2.6a) and (2.6b). The state $|e\rangle$ is a linear combination of $|JM\rangle$ that excludes any contribution $|J_t M_t\rangle$ appearing together with $|J_t, -M_t\rangle$. Coefficients in $|e\rangle$ are $\Lambda(JM)$ and,

$$|e\rangle = \sum_{JM} |JM\rangle \Lambda(JM), \qquad (2.9)$$

with $\langle U^{K'}_{Q'}\rangle = \langle e|U^{K'}_{Q'}|e\rangle$ the corresponding expectation value of an Hermitian operator. A second state $|\bar{e}\rangle$ is constructed from (2.9) using a rule $\{|JM\rangle \Lambda(JM)\} \to \{(-1)^{J-M} |J, -M\rangle \Lambda(JM)^*\}$, namely,

$$|\bar{e}\rangle = \sum_{JM} (-1)^{J-M} |J, -M\rangle \Lambda(JM)^*. \qquad (2.10)$$

States $|e\rangle$ and $|\bar{e}\rangle$ have no common components, by their very construction, and are thus orthogonal, $\langle \bar{e}|e\rangle = 0$. Wavefunctions for a Kramers doublet can be represented by linear combinations of $|e\rangle$ and $|\bar{e}\rangle$. Application of the aforementioned rule to $|\bar{e}\rangle$ recovers $|e\rangle$ to within a sign that is $+1$ for an even number of electrons (all $J$ are integer) and $-1$ for and odd number of electrons (all $J$ are half-integer).

Identities (2.6) lead to,

$$\langle \bar{e}|U^{K'}_Q|\bar{e}\rangle = \sigma_\theta \, \sigma_\pi \langle e|U^{K'}_Q|e\rangle. \tag{2.11}$$

For a parity-even operator, with $\sigma_\pi = +1$, identity (2.11) confirms that $|\bar{e}\rangle$ can be used as the time reversed version of $|e\rangle$, i.e., a time-odd Hermitian operator obeys $\langle \bar{e}|U^{K'}_Q|\bar{e}\rangle = - \langle e|U^{K'}_Q|e\rangle$. Spin and orbital angular momentum operators are parity-even. Compound operators constructed with **s**, *l* or **J** are likewise parity-even, with time-odd compound operators made from an odd number.

The presence of spin, a relativistic quantity, requires consideration of the conjugate operators for time-reversal and space-inversion that reverse the sign of the four-vector $\{\mathbf{R}, t\}$. Accordingly, (2.11) applies in the general case and the quantity $[\sigma_\pi \langle \bar{e}|U^{K'}_Q|\bar{e}\rangle]$ is the time-reversed expectation value of an Hermitian operator $U^{K'}_{Q'}$ [6].

The expectation value of an anti-Hermitian operator $\check{U}^{K'}$ satisfies $\langle \check{U}^{K'}_{Q'}\rangle = -(-1)^{Q'} \langle \check{U}^{K'}_{-Q'}\rangle^*$. The corresponding RME obeys,

$$(Jsl\|\check{U}^{K'}\|J's'l')^* = -(-1)^{J-J'}(J's'l'\|\check{U}^{K'}\|Jsl), \tag{2.12}$$

which replaces (2.6a). The relation (2.6b) may apply to $\check{U}^{K'}$ but a counter example follows.

Consider an anti-Hermitian operator $\check{\mathbf{n}} = i\mathbf{n}$, where the electric dipole $\mathbf{n} = \mathbf{R}/R$ is parity-odd. The latter operator is Hermitian with an RME,

$$(l\|n\|l') = (-1)^l \sqrt{[(2l+1)(2l'+1)]} \begin{pmatrix} l & 1 & l' \\ 0 & 0 & 0 \end{pmatrix}, \tag{2.13}$$

that is purely real and vanishes unless $l + l'$ is odd and,

$$(jsl\|n\|j'sl') = (-1)^{s+l'-j}(l\|n\|l') \sqrt{[(2j+1)(2j'+1)]} \begin{Bmatrix} j & l & s \\ l' & j' & 1 \end{Bmatrix}, \tag{2.14}$$

obtained by direct calculation tells us that $(jsl\|\check{n}\|j'sl')$ is purely imaginary (in (2.14) the number $s - j$ in the phase factor is an integer). Using this fact in conjunction with $(l\|n\|l') = (-1)^{l-l'}(l'\|n\|l)$ from (2.13) shows that $(jsl\|\check{n}\|j'sl')$ derived from (2.14) does indeed obey (2.12). Alas, (2.6b) cannot be used for $\check{\mathbf{n}} = i\mathbf{n}$, because it is anti-Hermitian. We are obliged to define $\check{\mathbf{n}}$ as time-odd, whereas use of (2.14) in (2.6b) would have $\check{\mathbf{n}}$ time-even, given it is parity-odd and $K' = 1$.

The reason that identity (2.6b) does not work for $\check{\mathbf{n}}$ is a direct consequence of our definition of the time-signature,

$$\{\theta V^{K'} \theta^{-1}\}^\dagger = \sigma_\theta V^{K'}, \tag{2.15}$$

where $^\dagger$ denotes Hermitian conjugation and the time-reversal operator includes complex conjugation. Spin and angular momentum operators are time-odd, of course. Definition (2.15) preserves this property for their raising an lowering operators, e.g., $(S_x + iS_y)$ satisfies (2.15) with $\sigma_\theta = -1$, which is one reason why (2.15) is the conventional definition of the time signature $\sigma_\theta = \pm 1$ in studies involving magnetism. But, definition (2.15) applied to the anti-Hermitian operator $\check{\mathbf{n}}$ returns a time-even signature and this is not what we need in **D** defined through (1.4).

Additional insight to properties of θ is gained from its action on a wavefunction ψ(**R**, t). First, θψ(**R**, t) = {ψ(**R**, −t)}* and, secondly, {ψ(**H**)}* = θψ(**H**) = ψ(−**H**) in the presence of a magnetic field **H**. From the second expression one can derive the identity [6],

$$\langle \psi_1 | V^{K'} | \psi_2 \rangle_{\mathbf{H}} = \langle \psi_2 | \{\theta V^{K'} \theta^{-1}\}^\dagger | \psi_1 \rangle_{-\mathbf{H}}. \quad (2.16)$$

Result (2.16) provides yet another reason for our definition of the time-signature (2.15), for the mean value of an operator obeys,

$$\langle V^{K'} \rangle_{\mathbf{H}} = \sigma_\theta \langle V^{K'} \rangle_{-\mathbf{H}}. \quad (2.17)$$

Thus a magnetic (time-odd, $\sigma_\theta = -1$) operator changes sign when the polarity of the magnetic field is reversed, and the property is independent of the rank of the operator. Our definition for θ acting on a state $|JM\rangle$ is,

$$\theta\{c\,|JM\rangle\} = c^*\,\{(-1)^{J-M} P_\pi^{-1}\,|J, -M\rangle\}, \quad (2.18)$$

where c is a classical number and $P_\pi$ is the parity operator, i.e., $P_\pi V^{K'} P_\pi^{-1} = \sigma_\pi V^{K'}$. The identity (2.6b) is a direct consequence of (2.18).

Since an operator with discrete symmetries $\sigma_\theta = -1$ and $\sigma_\pi = -1$ is magneto-electric, **ň** is an anti-Hermitian magneto-electric dipole operator. An Hermitian magneto-electric dipole **Ω** of interest possesses an RME,

$$(l\|\mathbf{\Omega}\|l') = [l(l+1) - l'(l'+1)]\,(l\|\text{ň}\|l'), \quad (2.19)$$

that is purely imaginary and symmetric with respect to an interchange of l and l', which are the opposite properties of $(l\|n\|l')$ defined in (2.13). The result (2.14) applies also to **Ω**. The RME $(jsl\|\mathbf{\Omega}\|j'sl')$ is purely imaginary, and (2.6) tells us that discrete symmetry signatures of **Ω** satisfy $\sigma_\theta \sigma_\pi = +1$, as anticipated.

## 3. Orbital contribution to magnetic scattering

It is natural to separate the linear momentum operator **p** = −iℏ∇ in (1.1) into its angular and radial components that we label by the letters a and r [4, 5]. Thereafter, it can be shown that [1, 10],

$$-(i/\hbar k^2)\exp(i\mathbf{R}_j \cdot \mathbf{k})\,(\mathbf{k} \times \mathbf{p}_j) = \sum_{K,K'} i^K \{C^K(\boldsymbol{\kappa}) \otimes O^{K'}(a)\}^1 + \sum_{K'} i^{K'} \{C^K(\boldsymbol{\kappa}) \otimes O^{K'}(r)\}^1, \quad (3.1)$$

where we employ tensor products defined by (2.2). Multipole operators $O^{K'}(a)$ and $O^{K'}(r)$ are defined in (3.2) and (3.12). It seems that a derivation of the result in (3.1) and related results spreads over many pages of a notebook, and great care has to be taken with phase factors. Likewise for derivations of (3.4) and (3.5). The corresponding derivation of the orbital-spin contribution to scattering, reported in §4, is simple by comparison.

*Angular part*;

$$O^{K'}(a) = -(2K+1)\sqrt{[3(2K'+1)]}\sum_{x,y}(-1)^{(1+K+x)/2}(j_x)(2x+1)(2y+1) \quad (3.2)$$

$$\begin{pmatrix}1 & K & x \\ 0 & 0 & 0\end{pmatrix}\begin{pmatrix}1 & x & y \\ 0 & 0 & 0\end{pmatrix}\begin{Bmatrix}1 & x & y \\ K' & 1 & 1\end{Bmatrix}\begin{Bmatrix}K & K' & 1 \\ 1 & 1 & x\end{Bmatrix}\{L \otimes C^y(\mathbf{n}) - (-1)^{K'+y} C^y(\mathbf{n}) \otimes L\}^{K'},$$

where **L** and **n** are operators for orbital angular momentum and the electric dipole, respectively. The radial integral is,

$$(j_x) = \int_0^\infty dR R^2 f_l(R) f_{l'}(R) \{j_x(kR)/kR\}, \qquad (3.3)$$

with $j_x(x)$ a spherical Bessel function, and $f_l(R)$ and $f_{l'}(R)$ radial parts of electron orbitals labelled by their angular momenta $l$ and $l'$. The rank $y$ is even for equivalent electrons in an atomic shell ($l = l'$) [1]. In this case $O^{K'}(a)$ is parity-even, $x$ is odd, the operator vanishes unless $K'$ is odd, and $x = K'$. If electrons possess orbital angular momenta that differ by an odd integer the variable $y$ is odd ($l + l'$ odd) [10]. It follows that $x$ is even and $K$ is odd. One finds $y = K = K'$ for $K'$ odd, while $x = K'$ for $K'$ even. The case $y$ odd is allowed when the magnetic ion occupies a site denied a centre of symmetry, and the tensor operator $O^{K'}(a)$ is parity-odd.

An alternative expression for the angular part is,

$$\sum_{K,K'} i^K \{C^K(\boldsymbol{\kappa}) \otimes O^{K'}(a)\}^1 = i\,\boldsymbol{\kappa} \times \mathbf{D}(a), \qquad (3.4)$$

with a parity-odd dipole,

$$\mathbf{D}(a) = (1/\sqrt{2}) \sum_{K,K'} i^{K+1} (2K+1)\sqrt{(2K'+1)} (j_K) \sum_y (-1)^{K'+y} (2y+1) \qquad (3.5)$$

$$\times \begin{pmatrix} 1 & K & y \\ 0 & 0 & 0 \end{pmatrix} \begin{Bmatrix} K & K' & 1 \\ 1 & 1 & y \end{Bmatrix} \{C^K(\boldsymbol{\kappa}) \otimes \{\mathbf{L} \otimes C^y(\mathbf{n}) - (-1)^{K'+y} C^y(\mathbf{n}) \otimes \mathbf{L}\}^{K'}\}^1.$$

The variable $y$ is an odd integer when electrons possess orbital angular momenta that differ by an odd integer ($l + l'$ odd).

Radial parts of the orbitals are identical for equivalent electrons, $f_l(R) = f_{l'}(R)$, and radial integrals $\langle j_n(k) \rangle$ are defined by (4.3). The dipole approximation for $\mathbf{D}(a)$ is,

$$\mathbf{D}(a) \approx -(i/2) (\mathbf{L} \times \boldsymbol{\kappa}) \langle j_0(k) + j_2(k) \rangle. \qquad (3.6)$$

The combination of radial integrals $\langle j_0(k) + j_2(k) \rangle$ approaches unity for $k \to 0$. Extensive tabulations of $\langle j_n(k) \rangle$ are available in reference [11], and the Figure 1 illustrates $\langle j_0(k) \rangle$ for a 3d-transition ion and an actinide ion.

For $y$ odd ($l + l'$ odd) it is useful to separate contributions to $\mathbf{D}(a)$ from multipoles with $K'$ even and odd. We find,

$$\mathbf{D}(a) = (1/2) \sum_{(\text{odd})K'} i^{K'} [(1/3)(2K'-1)(2K'+1)/(K'+1)]^{1/2}$$

$$\times [(K'+1)(j_{K'-1}) - K'(j_{K'+1})] \{C^{K'-1}(\boldsymbol{\kappa}) \otimes \{\mathbf{L} \otimes C^{K'}(\mathbf{n}) - C^{K'}(\mathbf{n}) \otimes \mathbf{L}\}^{K'}\}^1$$

$$+ \sum_{(\text{even})K'} i^{K'+1} \{C^{K'}(\boldsymbol{\kappa}) \otimes Z^{K'}(\mathbf{n})\}^1. \qquad (3.7)$$

In the second part we use,

$$Z^{K'}(\mathbf{n}) = (1/2\sqrt{3})[(2K'+1)/(K'(K'+1))]^{1/2} (j_{K'})$$

$$\times [(K'+1)t(K') C^{K'-1}(\mathbf{n}) - K' t(K'+1) C^{K'+1}(\mathbf{n})], \qquad (3.8)$$

with,

$$t(K') = [\{(l + l' + 1)^2 - (K')^2)\}\{(K')^2 - (l - l')^2\}]^{1/2}. \tag{3.9}$$

Values of $K'$ even exclude $K' = 0$, for the contribution is zero on account of the orthogonality of **L** and **n**. In (3.7) the quantity $[i^{K'+1} Z^{K'}(\mathbf{n})]$ with $K'$ even contains the anti-Hermtian operator $\mathbf{ň} = i\mathbf{n}$, which is taken to be time-odd, as well as parity-odd. Operators with these properties are discussed in §2.

The lowest-order electronic tensor in **D**(a) is an orbital anapole,

$$\mathbf{\Omega} = \mathbf{L} \times \mathbf{n} - \mathbf{n} \times \mathbf{L} = i[\mathbf{L}^2, \mathbf{n}]. \tag{3.10}$$

An immediate consequence of the second form of $\mathbf{\Omega}$ in (3.10) is the RME (2.19), because $\mathbf{L}^2|lm\rangle = l(l + 1)|lm\rangle$. At the prescribed level of approximation,

$$\mathbf{D}(a) \approx -(1/4)\,\mathbf{\Omega}\,[2\,(j_0) - (j_2)]. \tag{3.11}$$

Matrix elements of (3.11) can be different from zero if the states involved have orbital angular momenta that differ by an odd integer, and such states are forbidden in the presence of a centre of inversion symmetry. By contrast, matrix elements of **L** in (3.6) are diagonal with respect to orbital angular momentum. Note that the radial integral $(j_0)$ is not bounded in the limit $k \to 0$. Illustrative examples of $(j_0)$ are found in the Figure 1.

*Radial part*; with orbital angular momenta $l$ and $l'$ that differ by an odd integer. The operator in (3.1) is,

$$O^{K'}(r) = (1/2\sqrt{3})\,(g_{K'})\,[K'\,(K' + 1)\,(2K' + 1)]^{1/2}\,C^{K'}(\mathbf{n}), \tag{3.12}$$

and $K'$ is an odd integer. The radial integral is,

$$(g_{K'}) = (2K' + 1) \int_0^\infty dR R^2 [f_l(R)\,(d/dR)f_{l'}(R) - f_{l'}(R)\,(d/dR)f_l(R)]\,\{j_{K'}(kR)/Rk^2\}, \tag{3.13}$$

which vanishes if all orbitals possess the same radial wavefunction. Illustrations of $(g_1)$ appears in the Figure 1. From (3.12) we obtain $O^1(r) = (1/\sqrt{2})\,(g_1)\,\mathbf{n}$, and the corresponding contribution in (3.1) is $[(i(g_1)/2)\,(\boldsymbol{\kappa} \times \mathbf{ň})]$.

An alternative form of the radial part of (3.1) is,

$$\sum_{K'} i^{K'} \{C^{K'}(\boldsymbol{\kappa}) \otimes O^{K'}(r)\}^1 = i\,\boldsymbol{\kappa} \times \mathbf{D}(r), \tag{3.14}$$

with a parity-odd dipole,

$$\mathbf{D}(r) = (1/2\sqrt{3}) \sum_{K'} i^{K'} (g_{K'})\,[K'\,(2K' - 1)\,(2K' + 1)]^{1/2}\,\{C^{K'-1}(\boldsymbol{\kappa}) \otimes C^{K'}(\mathbf{n})\}^1. \tag{3.15}$$

The dipole approximation is,

$$\mathbf{D}(r) \approx (1/2)\,(g_1)\,\mathbf{ň}, \tag{3.16}$$

and (3.14) is exactly the corresponding result derived from (3.12), as it should be. Properties of the anti-Hermtian operator $\mathbf{ň} = i\mathbf{n}$ are discussed in §2.

## 4. Orbital-spin contribution to magnetic scattering

The orbital-spin contribution in the definition (1.1) for **Q** can be written,

$$\exp(i\mathbf{R} \bullet \mathbf{k})\, \mathbf{s} = \sum_{K,K'} i^{K'+1} \sqrt{(2K+1)}\, \{C^K(\boldsymbol{\kappa}) \otimes H^{K'}\}^1, \tag{4.1}$$

with an Hermitian operator that uses a tensor product (2.7),

$$H^{K'} = (-i)^{1+K+K'} \sqrt{[(2K+1)(2K'+1)/3]}\, (h_K)\, \{\mathbf{s} \otimes C^K(\mathbf{n})\}^{K'}. \tag{4.2}$$

The radial integral in (4.2) is,

$$(h_K) = \int_0^\infty dR\, R^2\, f_l(R)\, f_{l'}(R)\, j_K(kR). \tag{4.3}$$

Parity-even scattering by equivalent electrons is treated in §7, where the corresponding RME for $H^{K'}$ is related to a quantity $B(K, K')$ that is tabulated for nd- and nf-electrons. For the moment we note that $B(K' \pm 1, K')$ is a sum of two radial integrals $\langle j_n(k) \rangle$ with $n = K' \pm 1$, and $B(K, K) \propto \langle j_K(k) \rangle$. Radial integrals $\langle j_n(k) \rangle$ are defined by the right-hand side of (4.3) evaluated with $f_l(R) = f_{l'}(R)$, and the result $\langle j_0(0) \rangle = 1$ follows from $j_0(0) = 1$ and the normalization of $f_l(R)$. For parity-odd contributions to scattering we note for each $K$ there are three allowed values of $K'$. Radial integrals $(h_1)$ are illustrated in the Figure 1.

A dipole approximation for $l + l'$ odd follows on using $K = K' = 1$. One finds,

$$\exp(i\mathbf{R} \bullet \mathbf{k})\, \mathbf{s} \approx i\, (3/2)\, (h_1)\, [\boldsymbol{\kappa} \times (\mathbf{s} \times \mathbf{n})]. \tag{4.4}$$

The operator $(\mathbf{s} \times \mathbf{n})$ for the spin anapole was studied by Zel'dovich in the course of investigating parity-violating interactions in electromagnetic theory [14]. For $K = 1$ there are two values of $K'$, and the foregoing does not include the quadrupole contribution $K' = 2$. However, the quadrupole is included in both (4.5) and results (6.5) - (6.7) for $\mathbf{Q}_\perp$.

An alternative derivation of (4.4) is enlightening, in part because it naturally includes the absent quadrupole. If we treat $\mathbf{k}$ as a small quantity, $[\exp(i\mathbf{R} \bullet \mathbf{k})\mathbf{s}] \approx \{\mathbf{s} + (i\mathbf{R} \bullet \mathbf{k})\mathbf{s}\}$ where the leading term appears in Schwinger's result [12]. Let us re-write the manifestly parity-odd correction;

$$(i\mathbf{R} \bullet \mathbf{k})s_\alpha = \sum_\beta i k_\beta\, s_\alpha R_\beta = (kR) \sum_\beta i\kappa_\beta \{(1/3)\delta_{\alpha\beta}\, \mathbf{s} \bullet \mathbf{n} + (1/2) \sum_\gamma \varepsilon_{\alpha\beta\gamma}\, (\mathbf{s} \times \mathbf{n})_\gamma$$

$$+ (1/2)\, [s_\alpha n_\beta + s_\beta n_\alpha - (2/3)\, \delta_{\alpha\beta}\, \mathbf{s} \bullet \mathbf{n}]\}. \tag{4.5}$$

The structure of the identity (4.5), the sum of a scalar, a dipole and a quadrupole created from $\mathbf{s}$ and $\mathbf{n}$, can be understood from application of the triangle rule to the product of two dipoles, $\mathbf{n} = \mathbf{R}/R$ and $\mathbf{s}$. The scalar $(\mathbf{s} \bullet \mathbf{n})$ does not contribute to $\mathbf{Q}_\perp$, because it is multiplied by $\boldsymbol{\kappa}$. The dipole $\sum \kappa_\beta \varepsilon_{\alpha\beta\gamma}\, (\mathbf{s} \times \mathbf{n})_\gamma = [\boldsymbol{\kappa} \times (\mathbf{s} \times \mathbf{n})]_\alpha$ yields (4.4). Our definition of the quadrupole in the final contribution to (4.5) is standard and the trace is zero. Note that the anapole and quadrupole contributions in (4.5) are of the same order in $j_1(kR) \approx (kR)/3$.

## 5. Dipole approximation for parity-even scattering [1, 12, 13]

We record an approximate expression for the parity-even dipole that is frequently used to interpret elastic (diffraction) and inelastic scattering data. Its widespread use stems from both its simplicity and ability to capture the essence of contributing events. The so-called dipole approximation to the parity-even dipole $\mathbf{T}^1$ is a sum of operators for spin and orbital angular momentum, $\mathbf{S}$ and $\mathbf{L}$. These operators in $\mathbf{T}^1$ are weighted by radial integrals $\langle j_0(k) \rangle$ and $\langle j_2(k) \rangle$ defined by (4.3). The dipole approximation is,

$$\mathbf{T}^1 \approx (1/3) \{2 \, \mathbf{S} \, \langle j_0(k) \rangle + \mathbf{L} \, [\langle j_0(k) \rangle + \langle j_2(k) \rangle]\}. \tag{5.1}$$

For 3d-ions it is often adequate to use $\mathbf{L} = (g - 2)\mathbf{S}$ where the gyromagnetic ratio $g \approx 2$, because orbital angular momentum is almost quenched. For rare-earth ions, by contrast, a strong spin-orbit coupling combines $\mathbf{S}$ and $\mathbf{L}$ to give various $\mathbf{J}$ values. The Landé factor $g_o$ is defined by $g_o \mathbf{J} = 2\,\mathbf{S} + \mathbf{L}$.

## 6. Unit-cell structure factors

We define a unit-cell electronic structure factor,

$$\Psi^K_Q = \sum_{\mathbf{d}} \exp(i \mathbf{d} \cdot \mathbf{k}) \, \langle U^K_Q \rangle_{\mathbf{d}}, \tag{6.1}$$

where sites labelled by vectors $\mathbf{d}$ in a cell are occupied by magnetic ions. The structure factor $\Psi^K_Q$ complies with all elements of symmetry in a magnetic space-group, because translation symmetry is used to relate environments of ions. Bulk properties of a material are described by $\Psi^K_Q$ evaluated for $\mathbf{k} = 0$ and they are prescribed by elements of symmetry in the crystal class (point group). Additional physical properties of the material are allowed by the magnetic space-group and may contribute to Bragg spot intensities.

An expression for the intermediate scattering amplitude is conveniently derived using $\Psi^K_{\pm Q} = A^K_Q \pm B^K_Q$ for parity-even multipoles in magnetic neutron diffraction, i.e., $\langle U^K_Q \rangle = \langle T^K_Q \rangle$ in (6.1), and we add a superscript (+) to $\langle \mathbf{Q} \rangle$ to denote the parity signature $\sigma_\pi = +1$ of contributing multipoles. A unit vector $\boldsymbol{\kappa} = \mathbf{k}/k$ defines the direction of the Bragg wavevector in orthogonal coordinates (x, y, z). With the chosen notation $\Psi^1_x = -\sqrt{2}\, B^1_1$, $\Psi^1_y = i\sqrt{2}\, A^1_1$ and $\Psi^1_z = A^1_0$. The following expressions for $\langle \mathbf{Q} \rangle$ are derived by replacing $\langle T^K_Q \rangle$ by $\Psi^K_Q = A^K_Q + B^K_Q$ in either the general results (3.1) and (4.1), or (7.1) that is specific for parity-even equivalent electrons. Retaining dipoles ($K = 1$), quadrupoles ($K = 2$) and octupoles ($K = 3$) we find,

$$\langle Q_x \rangle^{(+)} \approx -(3/\sqrt{2})\, B^1_1 + \sqrt{3}\{\kappa_y \kappa_z [A^2_2 + \sqrt{(3/2)}\, A^2_0] + \kappa_x [i\kappa_z B^2_2 - \kappa_y B^2_1] + i(\kappa_y^2 - \kappa_z^2)\, A^2_1\}$$

$$+ (3/4)\sqrt{35}\{\kappa_x \kappa_z [\sqrt{(2/3)}\, A^3_2 - \sqrt{(1/5)}\, A^3_0] - \sqrt{(1/15)}\,(3\kappa_z^2 - 1)\, B^3_1 \tag{6.2}$$

$$+ i\kappa_x \kappa_y [A^3_3 - \sqrt{(1/15)}\, A^3_1] - i\sqrt{(2/3)}\, \kappa_y \kappa_z B^3_2 + (1/2)(\kappa_x^2 - \kappa_y^2)[\sqrt{(1/15)}\, B^3_1 - B^3_3]\},$$

$$\langle Q_y \rangle^{(+)} \approx (3i/\sqrt{2})\, A^1_1 + \sqrt{3}\{\kappa_x \kappa_z [A^2_2 - \sqrt{(3/2)}\, A^2_0] - i\kappa_y [\kappa_x A^2_1 + \kappa_z B^2_2] + (\kappa_x^2 - \kappa_z^2)\, B^2_1\}$$

$$+ (3/4)\sqrt{35}\{-\kappa_y \kappa_z [\sqrt{(2/3)}\, A^3_2 + \sqrt{(1/5)}\, A^3_0] + i\sqrt{(1/15)}\,(3\kappa_z^2 - 1)\, A^3_1 \tag{6.3}$$

$$+ \kappa_x \kappa_y [B^3_3 + \sqrt{(1/15)}\, B^3_1] - i\sqrt{(2/3)}\, \kappa_x \kappa_z B^3_2 + (i/2)(\kappa_x^2 - \kappa_y^2)[\sqrt{(1/15)}\, A^3_1 + A^3_3]\},$$

$$\langle Q_z \rangle^{(+)} \approx (3/2)\, A^1_0 + \sqrt{3}\{\kappa_z [i\kappa_x A^2_1 + \kappa_y B^2_1] - 2\kappa_x \kappa_y A^2_2 - i(\kappa_x^2 - \kappa_y^2)\, B^2_2\} \tag{6.4}$$

$$+ (1/4)\sqrt{7}\{(3/2)(3\kappa_z^2 - 1)\, A^3_0 + 4\sqrt{3}\,\kappa_z [i\kappa_y A^3_1 - \kappa_x B^3_1]$$

$$+ \sqrt{(15/2)}\,[(\kappa_x^2 - \kappa_y^2)\, A^3_2 - 2i\kappa_x \kappa_y B^3_2]\}.$$

The scattering amplitude for magnetic neutron Bragg diffraction $\langle \mathbf{Q}_\perp \rangle$ can be constructed using $\langle \mathbf{Q}_\perp \rangle = [\langle \mathbf{Q} \rangle - \boldsymbol{\kappa}\,(\boldsymbol{\kappa} \cdot \langle \mathbf{Q} \rangle)]$.

Universal expressions for Cartesian components of the intermediate amplitude for parity-odd multipoles, $\langle \mathbf{Q} \rangle^{(-)}$, at the chosen level of approximation, are simpler expressions than we encountered above for parity-even multipoles, because now we encounter quadrupoles rather than octupoles, in addition to dipoles. The relative simplicity of $\langle \mathbf{Q} \rangle^{(-)}$ prompts us to construct explicit expressions for the amplitude $\langle \mathbf{Q}_\perp \rangle^{(-)}$.

For the time-odd polar dipole (1.3) we use $\Psi^1 = \{\sum_\mathbf{d} \exp(i\mathbf{d} \cdot \mathbf{k}) \langle \mathbf{D} \rangle_\mathbf{d}\}$ with $\langle \mathbf{D} \rangle$ derived from (1.4). In addition, we retain quadrupoles $\langle \mathbf{H}^2 \rangle$ from (4.1) with $\langle \mathbf{H}^2 \rangle$ derived from (4.2) using $K = 1$ and $K' = 2$, because these multipoles are weighted by the radial integral $(h_1)$ that occurs with the spin anapole in $\langle \mathbf{D} \rangle$. Adopting a notation $\Psi^K_{\pm Q} = \mathcal{A}^K_Q \pm \mathcal{B}^K_Q$ the quadrupole contribution to $\langle \mathbf{Q} \rangle^{(-)}$ derived from (4.1) is simply $[-i\sqrt{3}\, \{C^1(\boldsymbol{\kappa}) \otimes (\mathcal{A}^2 + \mathcal{B}^2)\}^1]$. We go on to find,

$$\langle \mathbf{Q}_{\perp,x} \rangle^{(-)} \approx i[\kappa_y \Psi^1_z - \kappa_z \Psi^1_y] + (3/\sqrt{5})\, [i\kappa_x\{(2\kappa_y^2 + \kappa_z^2)\mathcal{A}^2_2 - \sqrt{(3/2)}\, \kappa_z^2 \mathcal{A}^2_0\}$$
$$+ (1 - 2\kappa_x^2)(\kappa_y \mathcal{B}^2_2 - i\kappa_z \mathcal{B}^2_1) + 2\kappa_x \kappa_y \kappa_z \mathcal{A}^2_1], \qquad (6.5)$$

$$\langle \mathbf{Q}_{\perp,y} \rangle^{(-)} \approx i[\kappa_z \Psi^1_x - \kappa_x \Psi^1_z] + (3/\sqrt{5})\, [-i\kappa_y\{(2\kappa_x^2 + \kappa_z^2)\mathcal{A}^2_2 + \sqrt{(3/2)}\, \kappa_z^2 \mathcal{A}^2_0\}$$
$$+ (1 - 2\kappa_y^2)(\kappa_x \mathcal{B}^2_2 - \kappa_z \mathcal{A}^2_1) + 2i\, \kappa_x \kappa_y \kappa_z \mathcal{B}^2_1], \qquad (6.6)$$

$$\langle \mathbf{Q}_{\perp,z} \rangle^{(-)} \approx i[\kappa_x \Psi^1_y - \kappa_y \Psi^1_x] + (3/\sqrt{5})\, [i\sqrt{(3/2)}\, \kappa_z (\kappa_x^2 + \kappa_y^2)\mathcal{A}^2_0 - i\kappa_z (\kappa_x^2 - \kappa_y^2)\mathcal{A}^2_2$$
$$- (1 - 2\kappa_z^2)(\kappa_y \mathcal{A}^2_1 + i\kappa_x \mathcal{B}^2_1) - 2\, \kappa_x \kappa_y \kappa_z \mathcal{B}^2_2]. \qquad (6.7)$$

Results (6.2) – (6.7) are used in §8.

### 7. Parity-even multipoles & equivalent electrons

A purely real quantity B($K$, $K'$) arises in matrix elements of $\sum \exp(i\mathbf{R}_j \cdot \mathbf{k})\mathbf{s}_j$ with a sum $j$ on equivalent electrons in an atomic shell with angular momentum $l$. Integers $K$ and $K'$ satisfy the triangle condition of a dipole, $K' = |K - 1|$, $K$, $K + 1$, and $K$ is an even integer. The condition on $K$ is imposed by the reduced matrix-element (RME) of $C^K(\mathbf{n})$ evaluated for equivalent electrons. A contribution $K = K'$ is allowed in the general B($K$, $K'$) although it is forbidden in many cases, as we shall see.

It is customary to express B($K$, $K'$) in terms of a unit tensor $W^{(a,K)K'}(\theta, \theta')$ with $\theta = J, S, L$ and $\theta' = J', S', L'$ (Russell-Saunders coupling scheme) as in (2.7), with $K = 0, 2, \ldots 2l$ and spin index $a = 1$ for the orbital-spin contribution under discussion. Selection rules for the unit tensor originate from the 9j-symbol it contains, c.f. (2.8),

$$\begin{Bmatrix} S & S' & a \\ L & L' & K \\ J & J' & K' \end{Bmatrix}.$$

An odd exchange of columns changes the sign of the 9j-symbol by a factor $(-1)^\mathfrak{R}$ with,

$$\mathfrak{R} = (S + S' + a + L + L' + K + J + J' + K').$$

If $\theta = \theta'$ two columns in the 9j-symbol are identical and it vanishes for $a + K + K'$ odd. This condition is not satisfied, however, when $K'$ is odd, for one has $K$ even for equivalent electrons and $a = 1$ in B($K$, $K'$). In this case, B($K$, $K'$) $\propto W^{(1,K)K'}(\theta, \theta)$ is related to a parity-even multipole that has a time-signature $\sigma_\theta = (-1)^{K'} = -1$ with $K' \leq (2l + 1)$, and a purely real RME. Should composite labels $\theta$ and $\theta'$ differ then $K = K'$ is allowed, the corresponding multipole is time-odd, of course, with $(\theta \| T^K \| \theta')$ purely imaginary. This attribute of $(\theta \| T^K \| \theta')$ is illustrated explicitly in subsequent work for a single electron.

The orbital contribution to neutron scattering $A(K, K')$, derived from $\sum \exp(i\mathbf{R}_j \cdot \mathbf{k}) (\mathbf{k} \times \mathbf{p}_j)$, has different properties from $B(K, K')$, although it is purely real also. Subsequent results are valid for an atomic shell. The condition $K'$ odd is deduced from (3.2), in which $y$ is even for equivalent electrons. Thus allowed values are $A(K' \pm 1, K')$ with $K' = 1, 3, ..., (2l - 1)$.

The intermediate operator with $K$ even and $K'$ odd can be written [1],

$$\mathbf{Q} = \sum_{K'} [(2K' + 1)/(K' + 1)] (2K' - 1)^{1/2} \{C^{K'-1}(\boldsymbol{\kappa}) \otimes \mathbf{T}^{K'}\}^{K'}$$

$$+ i \sum_K (2K + 1)^{1/2} \{C^K(\boldsymbol{\kappa}) \otimes \mathbf{T}^K\}^K. \qquad (7.1)$$

RMEs for the two parity-even and time-odd spherical tensors are,

$$(\theta \| \mathbf{T}^{K'} \| \theta') = -(-1)^{J'-J} (2J + 1)^{1/2} \{A(K' - 1, K') + B(K' - 1, K')\}, \qquad (7.2)$$

$$(\theta \| \mathbf{T}^K \| \theta') = -i(-1)^{J'-J} (2J + 1)^{1/2} B(K, K)$$

$$= (i^K/\sqrt{3}) (2K + 1) \langle j_K(k) \rangle (\theta \| \Upsilon^K(K) \| \theta'), \qquad (7.3)$$

and the operator $\Upsilon^K(K)$ is discussed later. Dependence of $(\theta \| \mathbf{T}^K \| \theta')$ on radial integrals is explicit in (7.3), while for $(\theta \| \mathbf{T}^{K'} \| \theta')$ radial integrals are included in $A(K' - 1, K')$ and $B(K' - 1, K')$, as we see in (7.15) and (7.17).

Absence in (7.1) and (7.2) of contributions $A(K' + 1, K')$ and $B(K' + 1, K')$ deserves comment. An intermediate operator of the form we define is arbitrary to within any function that is proportional to the scattering wavevector, $\mathbf{k}$, because the intermediate operator is related to the scattering operator $\mathbf{Q}_\perp$ through vector products with $\mathbf{k}$. We have exploited the arbitrariness to cancel contributions $K = (K' + 1)$, in the course of which use is made of the relation $(K' + 1)^{1/2} A(K' + 1, K') = (K')^{1/2} A(K' - 1, K')$ and a similar relation for $B(K, K')$ given explicitly in (7.13).

A tensor $\Upsilon^{I'}(I)$ defined as the tensor product of $\mathbf{s}$ and a spherical harmonic $C^I(\mathbf{n})$, where $\mathbf{n} = \mathbf{R}/R$ is the electric dipole moment, is useful in an investigation of properties of the orbital-spin contribution to the scattering amplitude that is not restricted to equivalent electrons - which explains momentary use of new integers $I$ and $I'$ for ranks of spherical tensors. Let,

$$\Upsilon^{I'}_Q(I) = (-i)^{I'+I+1} \sum_{pq} s_p C^I_q(\mathbf{n}) (1p\, Iq | I'Q), \qquad (7.4)$$

where $I' = |I - 1|, I, I + 1$. Because $s_p$ and $C^I_q(\mathbf{n})$ operate on different systems they commute. Inclusion of the phase factor in the tensor product (7.4) makes $\Upsilon^{I'}_Q$ an Hermitian operator,

$$[\Upsilon^{I'}_Q]^\dagger = (-1)^Q \Upsilon^{I'}_{-Q}. \qquad (7.5)$$

The condition $(I + l + l')$ even is enforced in $(\theta \| \Upsilon^{I'}(I) \| \theta')$ by the RME of $C^I(\mathbf{n})$. With quantum numbers $j, m, s, l$ ($s = 1/2$) identities (2.6) are,

$$(jsl\|\Upsilon^{I'}(I)\|j'sl')^* = (-1)^{j-j'}(j'sl'\|\Upsilon^{I'}(I)\|jsl) = \sigma_\theta \sigma_\pi (-1)^{I'}(jsl\|\Upsilon^{I'}(I)\|j'sl'). \quad (7.6)$$

Starting from (7.4) the second equality in (7.6) yields the relation $\sigma_\theta \sigma_\pi = (-1)^{I+1}$, which is interpreted as $\sigma_\theta = -1$, $\sigma_\pi = +1$ from the spin operator, and $\sigma_\theta = +1$, $\sigma_\pi = (-1)^I$ from $C^I(\mathbf{n})$. Returning to (4.1) and using the definition (7.4),

$$\exp(i\mathbf{R}\cdot\mathbf{k})s_p = \sum_I (2I+1)\langle j_I(k)\rangle \sum_{I'}[(2I'+1)/3]^{1/2} i^{I'+1}\{C^I(\boldsymbol{\kappa})\otimes\Upsilon^{I'}(I)\}^1_p. \quad (7.7)$$

This result is discussed for cases of particular interest.

The integer $I$ is even for equivalent electrons ($l = l'$), otherwise the RME of $C^I(\mathbf{n})$ is zero. For equivalent electrons we set $I = K$ even. We shall find that $(jsl\|\Upsilon^K(K)\|j'sl)$ is purely imaginary. Also, the result $(jsl\|\Upsilon^K(K)\|j'sl) = 0$ for $j = j'$ is a property of the unit tensor $W^{(1,K)K}(\theta,\theta')$ defined in (2.8). After some algebra we arrive at,

$$(jsl\|\Upsilon^K(K)\|j'sl) = (-i/2)(-1)^l[(2j+1)(2j'+1)/(2K+1)]^{1/2}(js\ j's|K\ 1). \quad (7.8)$$

The Clebsch-Gordan coefficient $(js\ j's|K\ 1) = 0$ for $j = j'$ ($K$ even) as anticipated, and $K$ has a minimum value 2. Setting $Q = 0$ in the definition (7.4) we may derive, for example,

$$\Upsilon^2_0(2) = \sqrt{(3/2)}\,(\mathbf{s}\times\mathbf{n})_z\, n_z, \qquad \Upsilon^4_0(4) = (\sqrt{5}/4)\,(\mathbf{s}\times\mathbf{n})_z\, n_z\,(7n_z^2 - 3). \quad (7.9)$$

In these two expressions, we recognize $(\mathbf{s}\times\mathbf{n})$ as the spin anapole [14], and $\Upsilon^2(2)$ and $\Upsilon^4(4)$ evidently obey $\sigma_\theta\sigma_\pi = -1$.

To complete a connection with previous work and tables for equivalent electrons ($l = l'$) [1, 3, 15] let,

$$C(K, K') = i^{K+K'+1}(-1)^{K'+j-j'}[(2K+1)/(2j+1)]^{1/2}\langle j_K(k)\rangle (jsl\|\Upsilon^{K'}(K)\|j'sl)$$

$$= (-1)^{K+K'+j-j'}[3(2K+1)/2(2j+1)]^{1/2}(l\|C^K(\mathbf{n})\|l)\langle j_K(k)\rangle W^{(1,K)K'}. \quad (7.10)$$

Note that $C(K, K')$ is purely real. One then has,

$$B(K, K) = i^K [(2K+1)/3]^{1/2} C(K, K); K\ \text{even}, \quad (7.11)$$

while,

$$B(K'-1, K') = i^{K'-1}[(K'+1)/(3(2K'+1))^{1/2}]$$

$$\times [C(K'-1, K') - (K'/(K'+1))^{1/2} C(K'+1, K')], \quad (7.12)$$

$$B(K'+1, K') = (K'/(K'+1))^{1/2} B(K'-1, K'); K'\ \text{odd}. \quad (7.13)$$

For the quantity $A(K'-1, K')$, which determines the orbital contribution for equivalent electrons in the reduced matrix-element $(\theta\|\mathbf{T}^{K'}\|\theta')$, we find,

$$A(K'-1, K') = i^{K'+1}(-1)^{j-j'}(2l+1)^2 [2(K'+1)/3(2j+1)]^{1/2}$$

$$\times \langle j_{K'-1}(k) + j_{K'+1}(k)\rangle A(K', K', l)\, W^{(0,K')K'}, \quad (7.14)$$

with $K' = 1, 3, ..., (2l-1)$. In this expression, $W^{(0,K')K'}$ appears because electron spin is absent and $a = 0$. A general expression for the quantity $A(K', K', l)$ is complicated [1], and we are content to give representative values; A(1, 1, 2) = 1/(5√5) and A(3, 3, 2) = − (√6/35√5). Of most practical use are actual values of A($K'$ − 1, $K'$), B($K$, $K$) and B($K'$ − 1, $K'$).

Entries in Table 1 for nd- and nf-electrons allow us to construct A($K'$ − 1, $K'$), B($K$, $K$) and B($K'$ − 1, $K'$). Expressions (7.2) and (7.3) relate these quantities to RMEs (reduced matrix-elements) of a parity-even spherical tensor operator (θ||**T**$^{K'}$||θ') and (θ||**T**$^K$||θ') with θ = $J$, $S$, $L$ and θ' = $J'$, $S'$, $L'$ (Russell-Saunders coupling scheme). We remind the reader that (θ||**T**$^{K'}$||θ') with $K'$ odd is purely real, while (θ||**T**$^K$||θ') with $K$ even is purely imaginary. RMEs (θ'||**T**$^K$||θ) and (θ'||**T**$^{K'}$||θ) are derived from tabulated quantities using (2.6a). Actual values in the tables are for fixed $S$ & $L$ specified in the standard notation, e.g., $^3$H for Pr$^{3+}$ (f $^2$) corresponds to $S$ = 1, $L$ = 5. Total angular momentum $J$ completes quantum numbers for the ground-state defined by Hund's rules, and we include various $J'$. Numbers in the tables are coefficients of radial integrals. We list A($K'$ − 1, $K'$)$_{K' − 1, K' + 1}$ in the expression,

$$A(K' - 1, K') = A(K' - 1, K')_{K' - 1, K' + 1} \langle j_{K' - 1}(k) + j_{K' + 1}(k) \rangle. \qquad (7.15)$$

For B($K$, $K$) and B($K'$ − 1, $K'$) we list B($K$, $K$)$_K$, B($K'$ − 1, $K'$)$_{K' − 1}$ and B($K'$ − 1, $K'$)$_{K' + 1}$ in expressions,

$$B(K, K) = B(K, K)_K \langle j_K(k) \rangle, \qquad (7.16)$$

$$B(K' - 1, K') = B(K' - 1, K')_{K' - 1} \langle j_{K' - 1}(k) \rangle + B(K' - 1, K')_{K' + 1} \langle j_{K' + 1}(k) \rangle. \qquad (7.17)$$

Blank spaces and zeroes in Table 1 are constructed from triangular rules applied to quantum numbers and ranks of tensors; an entry 0 is allowed by triangular rules but it is zero, whereas a blank entry is forbidden by triangular rules. As an example look at Pr$^{3+}$ (f $^2$) with A(2, 3)$_{2,4}$ zero for all $J'$, while for the same ion B(0, 1)$_0$ and B(0, 1)$_2$ are forbidden by triangular rules for $J'$ = 6.

## 8. Bragg diffraction by underdoped YBCO

A simulation of neutron diffraction by antiferromagnetic copper oxide shows that anapoles may contribute to intensities of Bragg spots [10]. The simulation exploits information extracted from resonant Bragg diffraction of soft x-rays [16]. In fact, a Cu wavefunction has been inferred from the soft x-ray data, and it is used to calculate the spin anapole. While neutron diffraction by anapoles in CuO remains to be tested, experiments on a cuprate superconductor already provide evidence that neutrons are deflected by Cu magneto-electric multipoles [17].

Intriguing results from two virtuoso experimental studies of the magnetic properties of the pseudo-gap phase of YBa$_2$Cu$_3$O$_{6 + x}$ (YBCO) can be interpreted by allowing Cu ions to possess a magneto-electric quadrupole. One experiment employed the Kerr effect [18, 19, 20] and the second Bragg diffraction of polarized neutrons [21, 22, 23]. As we shall see, both sets of results are compatible with magneto-electric quadrupoles at in-plane Cu sites in a ferro-type order using magnetic space-group Cm'm'm' depicted in Figure 2.

YBCO has a superconducting transition temperatures above that of boiling liquid nitrogen. It crystallizes in an orthorhombic-type structure with cell dimensions a ≈ 3.85 Å, c ≈ 11.7 Å [24]. A unit cell (Figure 3) contains three pseudo-cubic elementary perovskite unit cells and there are two CuO$_2$ plaquettes in a unit cell approximately 3.2 Å apart. The relation between oxygen concentration x and hole doping is non-trivial in YBCO, because the unit-cell contains two Cu types. Electronic properties are almost 2-dimensional and display strong angular anisotropy.[25] A gap associated with an insulating phase only exists for electrons travelling parallel to Cu-O bonds, whilst electrons travelling at 45° to this bond can move freely throughout the crystal. Low energy electrons reside in a single band of carriers formed by hybridization of $Cu^{2+}$ orbitals and oxygen 2p-electrons. An ordering wavevector consistent with a doubling of the in-plane chemical unit-cell describes the antiferromagnetic order in the $CuO_2$ planes of the parent compound. [26] This order disappears very fast when the material is doped.

The rotation of polarization of reflected light (Kerr effect) is a direct manifestation of broken time-reversal symmetry generally associated with long-range order in which there is a net magnetic moment. The space group proposed for the pseudo-gap phase of YBCO belongs to the magnetic crystal-class m'm'm' that allows the Kerr effect. In consequence, we are left to show that a ferro-type order of Cu magneto-electric quadrupoles also explains observed magnetic Bragg diffraction. A key finding from Bragg diffraction is that magnetic order in the pseudo-gap phase of YBCO is indexed on the chemical structure. For this reason polarization analysis was used to separate overlapping contributions to Bragg intensities from nuclear and magnetic scattering.

Axes for orthorhombic Cm'm'm' are derived from the parent tetragonal structure by rotation by 45°. Orthogonal (x, y, z) are {(0, 1, 0), (−1, 0, 0), (0, 0, 1)} and origin (0, 0, 0) with respect to the parent P4/mmm1'. Miller indices for Cm'm'm' are denoted (h, k, l) with $\kappa_x \propto h$, $\kappa_y \propto k$, and $\kappa_z \propto l$. One finds $h + k = 2H_o$, $h - k = -2K_o$ and $l = L_o$, where ($H_o$, $K_o$, $L_o$) are Miller indices for the tetragonal structure. We use $\kappa_a = k_a/k \propto H_o$, $\kappa_b = k_b/k \propto K_o$, $\kappa_c = k_c/k \propto L_o$ to define a unit vector $\mathbf{k}/k = (\kappa_a, \kappa_b, \kappa_c)$. It follows that $\langle \mathbf{Q}_a \rangle = (1/\sqrt{2})(\langle \mathbf{Q}_y \rangle + \langle \mathbf{Q}_x \rangle)$, $\langle \mathbf{Q}_b \rangle = (1/\sqrt{2})(\langle \mathbf{Q}_y \rangle - \langle \mathbf{Q}_x \rangle)$ and $\langle \mathbf{Q}_c \rangle = \langle \mathbf{Q}_z \rangle$ in the intermediate amplitude ($\langle \mathbf{Q}_a \rangle$, $\langle \mathbf{Q}_b \rangle$, $\langle \mathbf{Q}_c \rangle$).

The experiment measured Bragg intensities in spin-flip (SF) = $|\langle \mathbf{Q}_\perp \rangle|^2 - |\mathbf{P} \cdot \langle \mathbf{Q}_\perp \rangle|^2$, with primary polarization of the neutron beam $\mathbf{P}$ a unit vector. No intensity could be observed in the Bragg spot (0, 0, 2) with $\mathbf{P} // \mathbf{k}$. However, intensity was observed in the Bragg spot (1, 0, 1) for three configurations of $\mathbf{P}$, and we label them (a), (b) and (c). One has $\mathbf{k}/k = (\kappa_a, 0, \kappa_c)$ with $(\kappa_a^2 + \kappa_c^2) = 1$ and $\kappa_a \approx 0.950$ at (1, 0, 1) and,

(a) $\mathbf{P} // \mathbf{k}$, SF = $|\langle \mathbf{Q}_\perp \rangle|^2$.

(b) $\mathbf{P} \cdot \mathbf{k} = 0$ using $\mathbf{P} = (-\kappa_c, 0, \kappa_a)$, and $\mathbf{P} \cdot \langle \mathbf{Q}_\perp \rangle = (\kappa_a \langle \mathbf{Q}_c \rangle - \kappa_c \langle \mathbf{Q}_a \rangle)$ leads to SF = $|\langle \mathbf{Q}_b \rangle|^2$.

(c) $\mathbf{P} \cdot \mathbf{k} = 0$ using $\mathbf{P} = (0, 1, 0)$, and $\mathbf{P} \cdot \langle \mathbf{Q}_\perp \rangle = \langle \mathbf{Q}_b \rangle$ leads to SF = $|\kappa_a \langle \mathbf{Q}_c \rangle - \kappa_c \langle \mathbf{Q}_a \rangle|^2$. (8.1)

Data sets are not independent since SF(a) = SF(b) + SF(c).

Copper ions in the magnetic space-group Cm'm'm' use sites 4k that have site symmetry m'm'2. The site symmetry requires a multipole $\langle U^K_Q \rangle$ to be unchanged by three operations, namely, $C_{2z}$, $m_x$' and $m_y$', where $m_x = IC_{2x}$ is a mirror on the x-axis and a prime denotes time reversal (imposition of symmetry operations on multipoles is discussed at length by Lovesey and Balcar [27]). Symmetry with respect to $C_{2z}$ is satisfied by $Q$ even. In this case $m_x' \equiv m_y'$ with,

$$\langle U^K_Q \rangle = IC'_{2x} \langle U^K_Q \rangle = \sigma_\pi \sigma_\theta (-1)^K \langle U^K_{-Q} \rangle. \tag{8.2}$$

Magnetic, parity-even multipoles denoted by $\langle T^K_Q \rangle$ possess $\sigma_\pi \sigma_\theta = -1$. Thus, a magnetic dipole is allowed along the z-axis, and there is one component of the quadrupole that is purely imaginary, $\langle T^2_2 \rangle = -\langle T^2_{-2} \rangle = -\langle T^2_2 \rangle^*$. Magneto-electric multipoles possess $\sigma_\pi \sigma_\theta = +1$, and an anapole ($K = 1$) is forbidden while a purely real quadrupole $\langle H^2_Q \rangle$ with $Q = 0, \pm 2$ is allowed. Turning to the electronic structure factor (6.1) we find,

$$\Psi^K_Q = [1 + (-1)^{h+k}] [\exp(i\varphi) + \sigma_\pi \sigma_\theta \exp(-i\varphi)] \langle U^K_Q \rangle, \tag{8.3}$$

with $\varphi = 2\pi lz$ and position parameter $z \approx 0.36$. Since $\Psi^K_Q \propto \langle U^K_Q \rangle$ the electronic structure factor obeys (8.2). The first factor in (8.3) follows from the centring condition, and $h + k = 2H_o$. Environments of ions in a cell are related by $C_{2x}$ whose effect on a multipole is nothing more than multiplication by the factor $\sigma_\pi \sigma_\theta$, as a consequence of the site symmetry (8.1). Setting $h = k = l = 0$ in (8.2) we find $\Psi^K_Q = 0$ for $\sigma_\pi \sigma_\theta = -1$, which is correct for a simple antiferromagnetic motif of magnetic moments.

Available experimental data imply that underdoped YBCO does not possess magnetic dipole moments [25, 28]. Let us, therefore, consider diffraction by parity-even quadrupoles, and use (6.2) – (6.3). Since $\Psi^2_Q = -\Psi^2_{-Q}$ we are led to $A^2_0 = A^2_2 = 0$ while $B^2_2 = \Psi^2_2 \propto \langle T^2_2 \rangle''$ is purely real. One finds,

$$\langle \mathbf{Q}_x \rangle^{(+)} \approx i\sqrt{3} \kappa_x \kappa_z B^2_2, \quad \langle \mathbf{Q}_y \rangle^{(+)} \approx -i\sqrt{3} \kappa_y \kappa_z B^2_2, \text{ and } \langle \mathbf{Q}_z \rangle^{(+)} \approx -i\sqrt{3}(\kappa_x^2 - \kappa_y^2) B^2_2. \tag{8.4}$$

With $\kappa_x = \kappa_y = 0$ for (0, 0, 2) the Bragg spot has no intensity in accord with the measurement. However, at the Bragg spot (1, 0, 1) where $\kappa_x = \kappa_y$ the result SF(c) = 0 derived from (8.1) does not agree with measurements, and parity-even quadrupoles as an explanation of neutron diffraction data can be discarded.

Turning to parity-odd multipoles in search of an explanation we find allowed quadrupoles $\langle H^2_Q \rangle$ with $Q = 0, \pm 2$ create a scattering amplitude that is fully consistent with observations. Using $\sigma_\pi \sigma_\theta = +1$ in (8.3) we find $\mathcal{B}^2_Q = 0$. Allowed $\mathcal{A}^2_Q$ are $\mathcal{A}^2_0 = \Psi^2_0$ and $\mathcal{A}^2_2 = \Psi^2_{+2}$, with $\Psi^2_Q = 4\cos(\varphi) \langle H^2_Q \rangle'$, from which we get,

$$\langle \mathbf{Q}_x \rangle^{(-)} \approx -(3i/\sqrt{5}) \kappa_x [(1/\sqrt{6}) \mathcal{A}^2_0 - \mathcal{A}^2_2], \quad \langle \mathbf{Q}_y \rangle^{(-)} \approx -(3i/\sqrt{5}) \kappa_y [(1/\sqrt{6}) \mathcal{A}^2_0 + \mathcal{A}^2_2],$$

$$\langle \mathbf{Q}_z \rangle^{(-)} \approx (3i/\sqrt{5}) \kappa_z \sqrt{(2/3)} \mathcal{A}^2_0. \tag{8.5}$$

These amplitudes yield null intensity for the Bragg spot (0, 0, 2), as required. And all spin-flip (SF) intensities (8.1) are different from zero. With regard to the intensity of the (1, 0, 1) Bragg spot we note that the magnitude of the wavevector $k \approx 1.7$ Å$^{-1}$ almost matches the first maximum of the radial integral ($h_1$) displayed in Figure 1 for Cu$^{2+}$: 3d$^8$ - 4p$^1$. Thus, a ferro-type order of Cu magneto-electric quadrupoles using Cm'm'm' is consistent with available data for the pseudo-gap phase of YBCO gathered by magnetic neutron diffraction [21, 22, 23], and it allows the Kerr effect that has been observed [18, 19, 20].

## 9. Discussion

Working with an atomic picture of electron valence states in magnetic materials we have reviewed the theory of magnetic neutron scattering. We have taken it beyond the conventional theory, developed by Schwinger and Trammell [12, 13], by including events that are allowed in the absence of inversion symmetry at a local level. Best known examples of which are magnetic charge - not visible in neutron scattering but visible in x-ray diffraction - and anapoles, also called toroidal dipoles. Our theory makes use of irreducible tensors and their associated multipoles.

Parity-even multipoles are related to the existing work on neutron scattering. Tables of reduced matrix-elements are provided for the important case of equivalent electrons nd and nf. Parity-odd multipoles, called magneto-electric multipoles in keeping with current terminology, are discussed at length because they are unlikely to be familiar to researchers for whom neutron scattering is their principal experimental tool. By way of an example of scattering by a state of magnetic charge, we provide a strong case for magneto-electric quadrupoles in the pseudo-gap phase of a cuprate superconductor (YBCO) delivering Bragg diffraction.

**Acknowledgements** Tables were produced by Professor Ewald Balcar who contributed many helpful comments on a previous version of the communication. Professor G. van der Laan made atomic calculations shown in Figure 1, while Dr D D Khalyavin prepared Figures 2 & 3.

**References**
[1] Lovesey S W 1987 *Theory of neutron scattering from condensed matter* volume **2** (Oxford: Clarendon Press)

[2] Kittel C 1987 *Quantum theory of solids* (Hoboken, NJ: Wiley)

[3] Balcar E and Lovesey S W 1989 *Theory of magnetic neutron and photon scattering* (Oxford: Clarendon Press)

[4] Edmonds A R 1960 *Angular momentum in quantum mechanics* (Princeton, NJ: University Press)

[5] Varshalovich D A, Moskalev A N and Khersonskii V K 1988 *Quantum theory of angular momentum* (World Scientific: Singapore)

[6] Balcar E and Lovesey S W 2009 *Introduction to the graphical theory of angular momentum* Springer tracts in modern physics vol. 234 (Heidelberg: Springer)

[7] Cricchio F *et al* 2009 *Phys. Rev. Lett*. **103** 107202

[8] Spaldin N A *et al* 2013 *Phys. Rev*. B**88** 094429

[9] Joly Y http://www.neel.cnrs.fr./fdmnes

[10] Lovesey S W 2014 *J. Phys.*: *Condens. Matter* **26** 356001

[11] Brown P J volume C 2004 *International tables of crystallography* (The Netherlands: Springer)

[12] Schwinger J S 1937 *Phys. Rev*. **51** 544

[13] Trammell G T 1953 *Phys. Rev*. **92** 1387

[14] Zel'dovich Y B 1958 *JETP* **6** 1184

[15] Lovesey S W *et al* 2005 *Physics Reports* **411** 233

[16] Scagnoli V *et al* 2011 *Science* **332** 696

[17]  Lovesey S W, Khalyavin D D and Staub U 2015 arXiv 1501.02427

[18] Xia J *et al* 2008 *Phys. Rev. Lett*. **100** 127002

[19] Kapitulnik A *et al* 2009 *New J. Phys*. **11** 055060

[20] Orenstein J 2011 *Phys. Rev. Lett*. **107** 067002

[21] Fauqué B *et al* 2006 *Phys. Rev. Lett*. **96** 197001

[22] Bourges P and Sidis Y 2011 *Comptes Rendus Physique* **12** 461

[23] Mangin-Thro L *et al* 2015 arXiv 1501.04919

[24] Wu M K *et al* 1987 *Phys. Rev. Lett*. **58** 908

[25] Keimer B *et al* 2015 *Nature* **518** 179

[26] Tranquada J M *et al* 1988 *Phys. Rev. Lett*. **60** 156

[27] Lovesey S W and Balcar E 2013 *J. Phys. Soc. Japan* **82** 021008

[28] Wu T *et al* 2014 arXiv 1404.1617

**Table 1**. Quanities defined in equations (7.15). (7.16) and (7.17) from which RMEs (7.2) and (7.3) are derived are listed for equivalent atomic electrons in shells nd and nf. Russell-Saunders coupling and Hund's rule determine ground state values of the quantum numbers $S$, $L$, $J$. Various values of $J'$ are included. Blank spaces and zeroes are constructed from triangular rules applied to quantum numbers and ranks of tensors; an entry 0 is allowed by triangular rules but it is zero, whereas a blank entry is forbidden by triangular rules.

**Figure 2**. Ferro-type ordering of magneto-electric quadrupoles in $CuO_2$ planes for YBCO in the pseudo-gap phase using magnetic space-group Cm'm'm'. Scattering amplitude is given by (8.5) with $A^2_Q = 4\cos(\varphi) \langle H^2_Q \rangle'$. Arrows indicate spin directions in magneto-electric quadrupoles $\langle H^2_0 \rangle' \propto \langle 3S_z n_z - \mathbf{S} \cdot \mathbf{n} \rangle$ and $\langle H^2_{+2} \rangle' \propto \langle S_x n_x - S_y n_y \rangle$, together with their response to spatial or time inversion. Reproduced from reference [17].

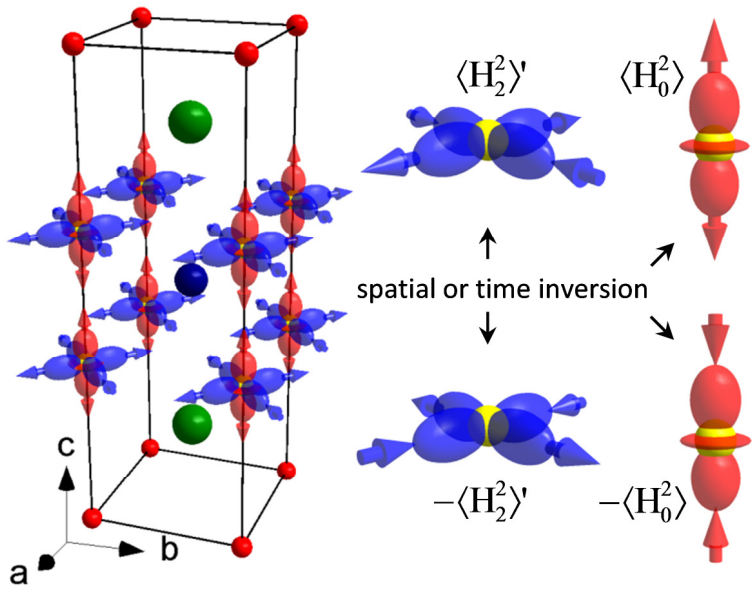

Figure 3. Crystal structure of YBaCu$_3$O$_7$ with the magnetic Cu2 ions in the CuO$_2$ planes. Reproduced from reference [17].

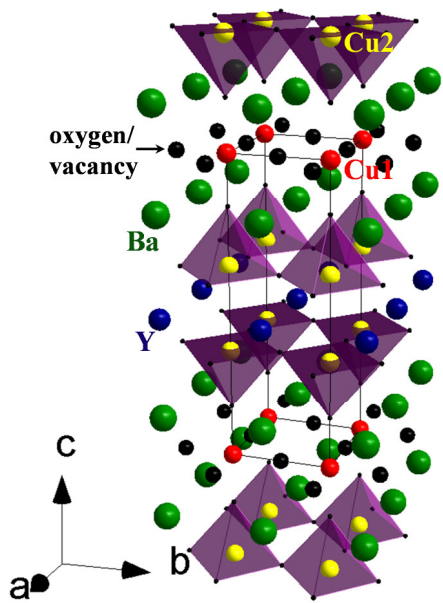

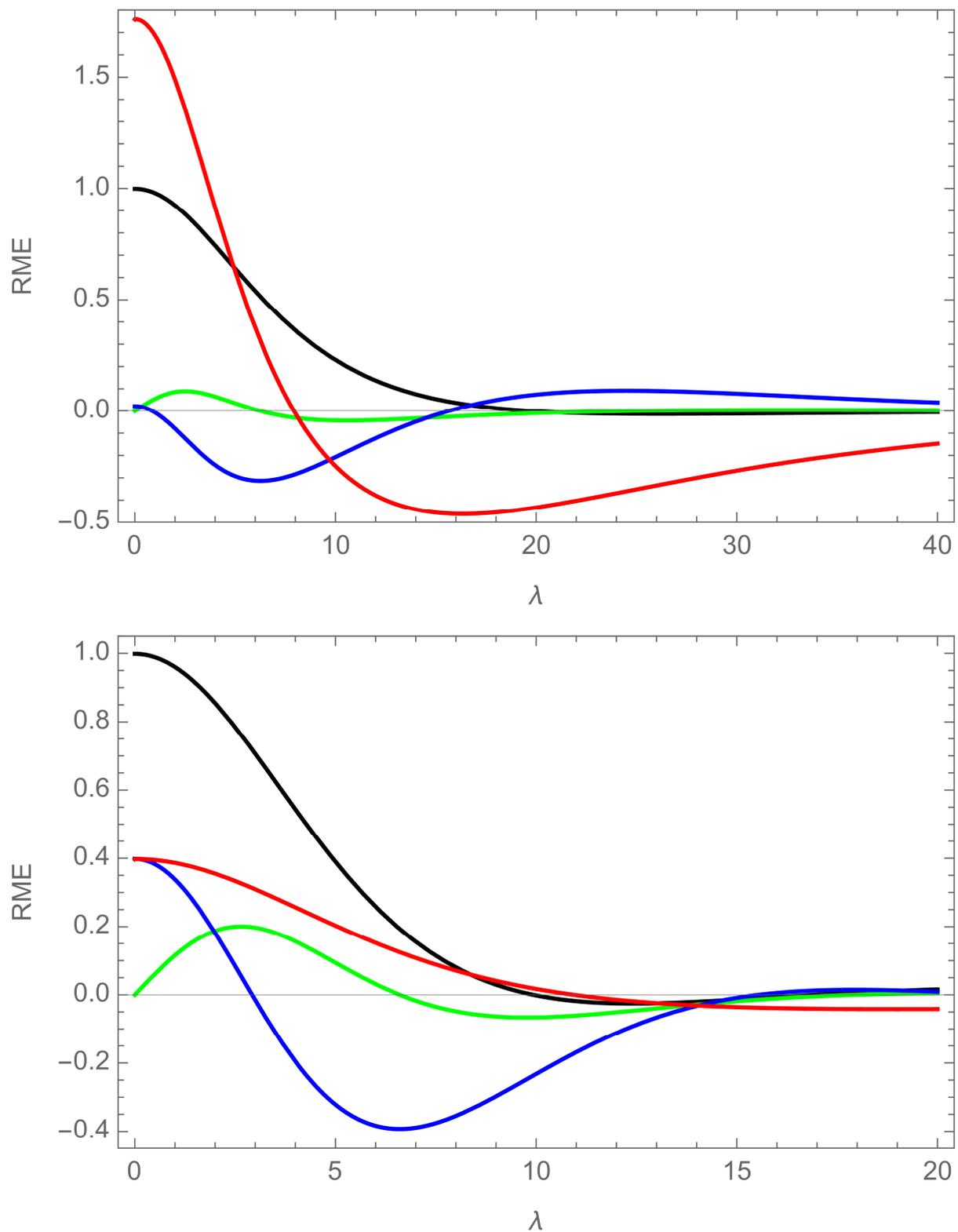

**Figure 1**. Radial integrals derived from an atomic code due to R Cowan. (———) $\langle j_0(k) \rangle$ defined by (4.3) for equivalent electrons $f_l(R) = f_{l'}(R) = 3d^8$ or $5f^2$. In addition; (———) $(h_1)$ defined by (4.3); (———) $\lambda \, (j_0)$ defined by (3.3); (———) $\lambda \, (g_1)$ defined by (3.13). A dimensionless wavevector $\lambda = 3a_o k$, where $a_o$ is the Bohr radius and $k$ the magnitude of the scattering wavevector. The corresponding variable in reference [11] is $s = (k/4\pi)$, which gives $\lambda = 12\pi a_o s$. Upper panel $Cu^{2+}$: $3d^8 - 4p^1$, and lower panel $U^{4+}$: $5f^2 - 6d^1$ displays $\{\lambda \, (g_1)/10\}$ (Calculations by G van der Laan, 2015).